\documentclass[aps, prd, showpacs, superscriptaddress, ctexart, nofootinbib, twocolumn]{revtex4-1}
\usepackage{graphicx}
\usepackage{epsfig}
\usepackage{bm}
\usepackage{verbatim}
\usepackage{amsfonts}
\usepackage[latin1]{inputenc}
\usepackage{graphicx}
\usepackage{caption}
\usepackage{subcaption}
\usepackage{amssymb}
\usepackage{color}
\usepackage{float}
\usepackage{amsmath}
\usepackage{amsfonts}
\usepackage{dcolumn}
\usepackage{bm}
\usepackage{tikz}
\usetikzlibrary{decorations.pathmorphing}

\def\be{\begin{equation}}
\def\ee{\end{equation}}
\def\ba{\begin{eqnarray}}
\def\ea{\end{eqnarray}}

\begin{document}
\title{ Exotica ex nihilo: Traversable wormholes \& non-singular black holes \\ from the vacuum of quadratic gravity}
\author{Francis Duplessis}
\email[]{fdupless@asu.edu}
\affiliation{Department of Physics, Arizona State University, Tempe, AZ 85287}

\author{Damien A. Easson}
\email[]{easson@asu.edu}
\affiliation{Department of Physics  \& Beyond Center for Fundamental Concepts in Science,\\
Arizona State University, Tempe, AZ 85287}

\begin{abstract}
We present new traversable wormhole and non-singular black hole solutions in  pure, scale-free $R^2$ gravity. These exotic solutions require no null energy condition violating or ``exotic" matter and are supported only by the vacuum of the theory. It is well known that $f(R)$ theories of gravity may be recast as dual theories in the Einstein frame. The solutions we present are found when the conformal transformation required to  move to the dual frame is singular. For quadratic $R^2$ gravity, the required conformal factor is identically zero for spacetimes with $R=0$. Solutions in this case are argued to arise in the strong coupling limit of General Relativity. 
\end{abstract}

\maketitle
\section{Introduction}
Recently, an exploration of the vacuum solutions of pure $R^2$ gravity uncovered new black hole solutions, resulting, in part, from the lack of a generalized Birkhoff theorem \cite{Kehagias:2015ata}. As a theory, quadratic gravity has several intriguing features: it possesses scale invariance and with the addition of the Weyl curvature tensor $C^2=C_{\mu\nu\rho\sigma}C^{\mu\nu\rho\sigma}$, is renormalizable \cite{Stelle:1976gc}. Numerically it has been shown
that the most general quadratic gravitational action, $R^2-\alpha C^2$ for some constant $\alpha$, also admits new black hole solutions \cite{Lu:2015cqa}. Pure $R^2$ gravity is the only pure quadratic gravity that is ghost free \cite{Kounnas:2014gda, Alvarez-Gaume:2015rwa} and, as an example of an $f(R)$ theory, is sometimes dual to the Einstein-Hilbert action of General Relativity (GR) minimally coupled to a scalar field. We briefly review how this duality arises in 
section~\ref{sec_review} and discuss the cases where the duality fails. It is precisely in this regime, where $R=0$, that the new black hole solutions of \cite{Kehagias:2015ata} were found. The authors interpreted these solutions as a part of the strong coupling limit of GR. In this paper, we show that the vacuum space of $R^2$ gravity permits new traversable wormhole and non-singular black hole solutions.
We emphasize that these structures are supported only by the vacuum and do not require any unusual states of matter. 

In Einstein's theory of General Relativity, every metric $g_{\mu\nu}$, is a solution to Einstein's equations for some associated stress energy tensor  $T_{\mu\nu}$.
It is therefore a challenging task to determine which solutions should be considered physical solutions. This dilemma lead to the development of \it ad hoc \rm energy conditions intended to reasonably restrict properties of the sourcing matter. The weakest such condition, the Null Energy Condition (NEC), stipulates that the stress energy tensor of matter should satisfy $T_{\mu\nu}k^\mu k^\nu\ge 0$ for any arbitrary null vector $k^\mu$. Imposing such conditions prohibits the construction of many spacetimes in the context of GR, including traversable wormholes \cite{Morris:1988cz}.  Wormholes have appeared in many science fiction settings, recently giving rise to additional research on the subject \cite{James:2015ima}. Besides a means of rapid interstellar travel used by advanced civilizations, wormholes are discussed in the quest to understand the relation between entanglement and the possible emergence of spacetime \cite{VanRaamsdonk:2010pw}. Such ideas were applied to the information paradox, yielding the so called,  $ER=EPR$ conjecture \cite{Maldacena:2013xja}, which suggests that entangled particles are connected via a (non-traversable) Einstein-Rosen bridge. Likewise, it has long been speculated that quantum gravity may somehow resolve singularities in 
black hole spacetimes leading some researchers to the idea that the Universe itself, might have been created on the interior of a black hole (for some early work see \cite{FMM,Morgan,MB,TMB,TV,PBB,Lowe,DKH,Smolin,Easson:2001qf,Easson:2002tg}).

In section~\ref{sec_review} we discuss how the new spacetime solutions are possible and present our solutions in
section~\ref{sec:sol}. We then show explicitly in section~\ref{sec_geoobs} that observers and light rays can traverse the throat of the wormhole and discuss some of the trajectory's properties in section~\ref{sec_div}. Finally, we discuss the appearance of the wormhole to asymptotic observers in section~\ref{sec_asymptoticsol}.
\section{$R^2$ gravity and its vacuum structure}\label{sec_review}
The $R^2$ gravity action is a specific example of the more general $f(R)$ action \cite{DeFelice:2010aj},
\be\label{fraction}
S=\frac{M_p^2}{2}\int d^4x\sqrt{-g} f(R)+S_m.
\ee
When the matter action $S_m$ exhibits conformal symmetry and $F(R)=df/dR$ is non-vanishing, it is well known that this action is equivalent to Einstein gravity minimally coupled to a scalar field. This is shown by writing the action in terms of the metric $\tilde{g}_{\mu\nu}=\Omega^2 {g}_{\mu\nu}$, where the conformal factor is given by 
$\Omega=F(R)^{1/2}$. Therefore, any solutions found in such $f(R)$ theories will also be found in Einstein gravity coupled to a scalar field. However, as was shown in \cite{Kehagias:2015ata}, this is not necessarily the case when $F(R)$ vanishes, and the metric $\tilde{g}_{\mu\nu}$, is ill defined. Wormhole solutions in
$f(R)$ have been studied extensively~\cite{Hochberg:1990is,Furey:2004rq,Lobo:2009ip,Oliveira:2011vu,DeBenedictis:2012qz,Harko:2013yb}. 

In $f(R)$ gravity the metric ${g}_{\mu\nu}$ has equation of motion,
\begin{eqnarray}
F(R) R_{\mu\nu} -\frac{1}{2}f(R)g_{\mu\nu} - \nabla_\mu\nabla_\nu F(R)  \nonumber \\
+g_{\mu\nu}\Box F(R)=M_p^{-2} T_{\mu\nu}.
\end{eqnarray}
In the above, $T_{\mu\nu}$ is the stress-energy tensor of any external matter present.\footnote{We use the reduced Planck mass, $M_p^2=1/8\pi{G_N}$,  metric signature $(-,+,+,+)$,
and $\Box \equiv \nabla^\mu\nabla_\mu$.}  Focusing on vacuum solutions, $T_{\mu\nu}=0$, we notice that any metric producing $f(R)=F(R)=0$ will be a solution of this theory. These conditions may constraint $R$  but leave significant freedom in the form of the Ricci tensor $R_{\mu\nu}$.

Focusing on the simplest case which satisfies the above criteria, $f(R)=R^2$, we find that every spacetime with traceless Ricci tensor is a solution to the $R^2$ gravity EOM. This abundant freedom in the form of $R_{\mu\nu}$ is how Birkhoff's theorem is circumvented. In standard Einstein gravity, $R=0$ implies $R_{\mu\nu} \propto T_{\mu\nu}$ and therefore any metric sourced by matter with traceless energy-momentum tensor will also be a solution to the vacuum of $R^2$ gravity. One notable example is the electromagnetic field with $\mathcal{L}_m=-\frac{1}{4}F_{\mu\nu}F^{\mu\nu}$. 
We will show that this new freedom found in the vacuum of $R^2$ gravity allows bundles of ingoing modes to evolve from converging to diverging modes in section \ref{sec_div}. In
ordinary Einstein gravity, such evolution would only be possible through a violation of the NEC as can be shown through the Raychaudhuri equation--an equation which describes the evolution of the bundle's divergence $\theta$. For null geodesics with tangent null vector $k^\mu$, the Raychaudhuri equation is
\be
\frac{\text{d}}{\text{d}\lambda}\theta=-\frac{1}{2}\theta^2 -\sigma_{\mu\nu}\sigma^{\mu\nu}+\omega_{\mu\nu}\omega^{\mu\nu}-R_{\mu\nu}k^\mu k^\nu.
\ee
Here $\sigma_{\mu\nu}$ is the shear tensor and $\omega_{\mu\nu}$ the vorticity tensor. The vorticity can always be set to zero by choosing a coordinate system in which the congruences are hypersurface orthogonal. The remaining terms on the RHS are all negative in GR if the NEC is satisfied ($R_{\mu\nu}k^\mu k^\nu=M_P^{-2} T_{\mu\nu}k^\mu k^\nu\geq 0$). This implies that $\frac{\text{d}}{\text{d}\lambda}\theta<0$ and hence bundles of null geodesics are always converging. In $R^2$ gravity, vacuum solutions can have $R_{\mu\nu}k^\mu k^\nu>0$  so that the above conclusion need not apply. 
\section{Non-singular black holes and a Traversable Wormhole}\label{sec:sol}
The most general static and spherically symmetric spacetime is given by
\be
ds^2=-f_1(r)dt^2+f_2(r)dr^2+r^2d\Omega^2,
\ee
where $\text{d}\Omega^2=\text{d}\theta^2+\sin^2\!\theta \, \text{d}\phi^2$. We make the following metric ansatz
\be
 f_1(r)=G(r)\text{   ~~ and,   ~~~  }f_2(r)=\frac{r^2}{r^2-k^2}\frac{1}{G(r)}.
\ee
Introducing a reparametrization of the radial coordinate, $l^2=r^2-k^2$,  for some constant $k$, the metric becomes,
\be\label{ansatz}
\text{d}s^2=-G(l)\text{d}t^2+\frac{1}{G(l)}\text{d}l^2+(l^2+k^2)\text{d}\Omega^2.
\ee
If $G(l)\rightarrow 1$ as $l\rightarrow \pm \infty$, this metric connects two asymptotically Minkowski spacetimes. The constant $k$ will set the minimal wormhole throat radius as $r$ cannot become smaller than this value.

The Ricci scalar of this spacetime is given (in terms of $G$) by,
\begin{eqnarray}\label{ricciscalar}
-\frac{1}{2}(k^2+l^2)^2 R=\frac{1}{2}\Big((k^2+l^2)^2G'\Big)' \nonumber \\
+(k^2+l^2)(G-1)+G k^2 \,,
\end{eqnarray}
where  $'\equiv d/dl$.
Requiring $R=0$,  yields a two parameter solution for $G$. A complicated exact solution of this second order ODE is given in Appendix \ref{appA}. 
For our purposes we need only discuss certain limiting behaviors. Taking the large $l$ limit of Eq.(\ref{ricciscalar}), corresponding to large $r$, and solving the $R=0$ ODE, 
we find the asymptotic solution,
\be\label{largel}
G(l\rightarrow \pm \infty)\simeq 1-\frac{2M_{\pm}}{l}+\frac{Q_{\pm}^2}{l^2}.
\ee
This resembles a Reissner\textendash Nordstr{\"o}m metric of mass $M_{\pm}$ and charge $Q_{\pm}$. These need not be the same for the two different asymptotic regions. Hence we label them by a $\pm$ to denote the values measured by observers located at $l\rightarrow \pm\infty$. Note that these parameters arises as integration constants and should not be associated with a conserved ``mass" and ``charge" even though we denote them by $M$ and $Q$. For example, the approximate solution in equation (\ref{largel}) drops a non-analytic term, seen only in the large $l$ limit of the exact solution, that goes as $k^2\frac{\log(l/k)}{l^2}$. Hence $Q_\pm^2$, taken as the coefficient of the $1/l^2$ term in the series expansion, cannot be a well defined finite charge. We will discuss this point further in section \ref{sec_asymptoticsol}.

\begin{figure}[tbhp]
\centering
  \includegraphics[width=0.48\textwidth]{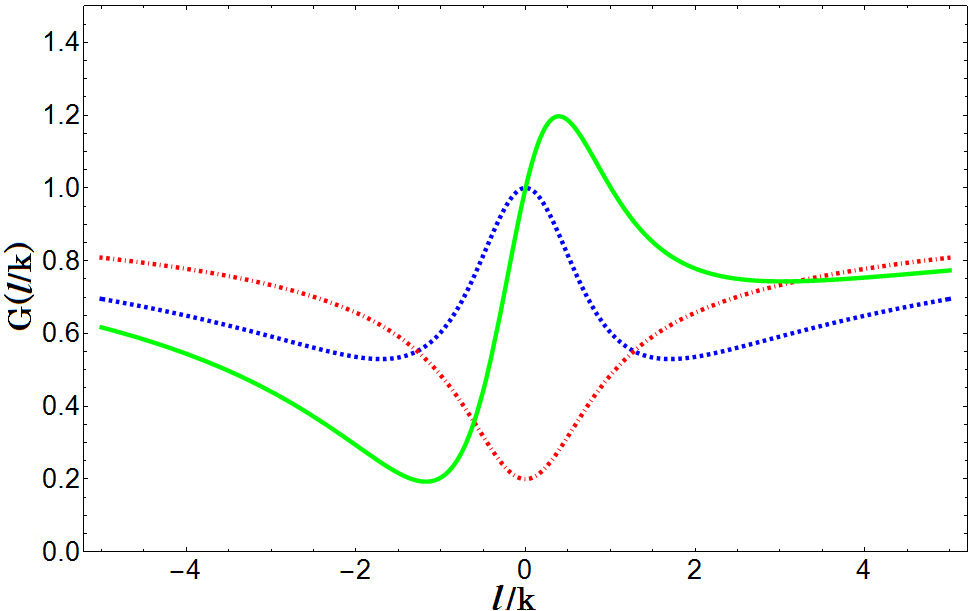}
  \caption{Plot of $G(l/k)$ vs $l/k$. The blue/dashed, red/dotted-dashed, and green/solid have initial conditions $(G_0,v_0)$ set as $(1,0),~(0.2,0),\text{and}~(1,1/k)$ respectively.}\label{worm1}
\end{figure}

At small $l$, the solution is approximately,
\be\label{smalll}
G(l\approx 0)\simeq \frac{1}{2}\big[1+ 2(G_0-0.5) \cos (2l/k)+kv_0 \sin (2l/k)\big]\,
\ee
where $G_0=G(0)$ and $v_0=G'(0)$ are the initial conditions determining the shape of the throat. The constants $M_\pm$ and $Q_\pm$ are expressed in terms of these parameters in section \ref{sec_asymptoticsol}. Given appropriate choices of $G_0,~v_0$, we ensure that $G$ is strictly positive for all $l$. The solutions are shown in figure (\ref{worm1}). In these cases the spacetime has no horizons even though it appears as a black hole to asymptotic observers.

\begin{figure}[tbhp]
 \centering
    \includegraphics[width=0.25\textwidth]{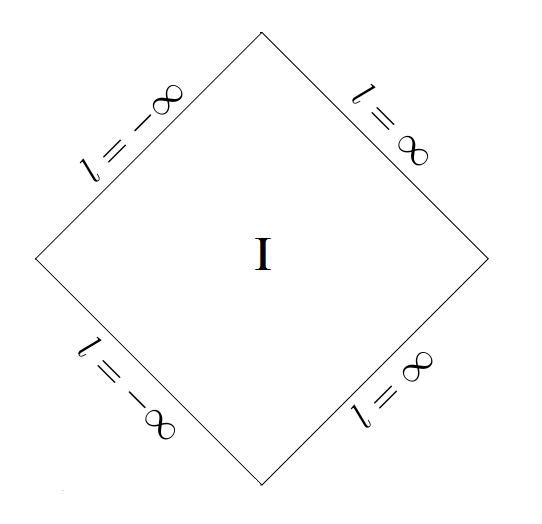}
  \includegraphics[width=0.40\textwidth]{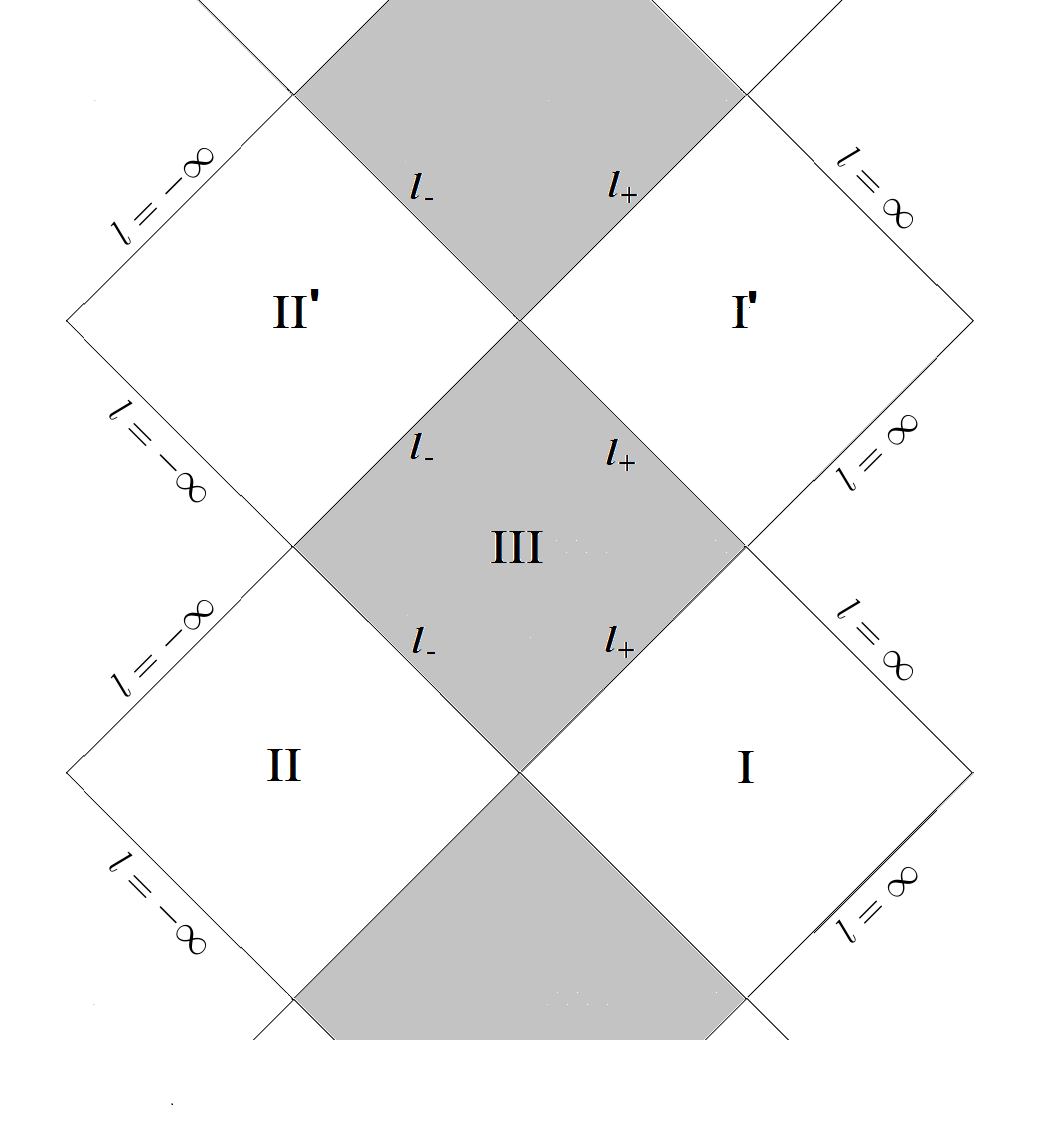}
  \includegraphics[width=0.48\textwidth]{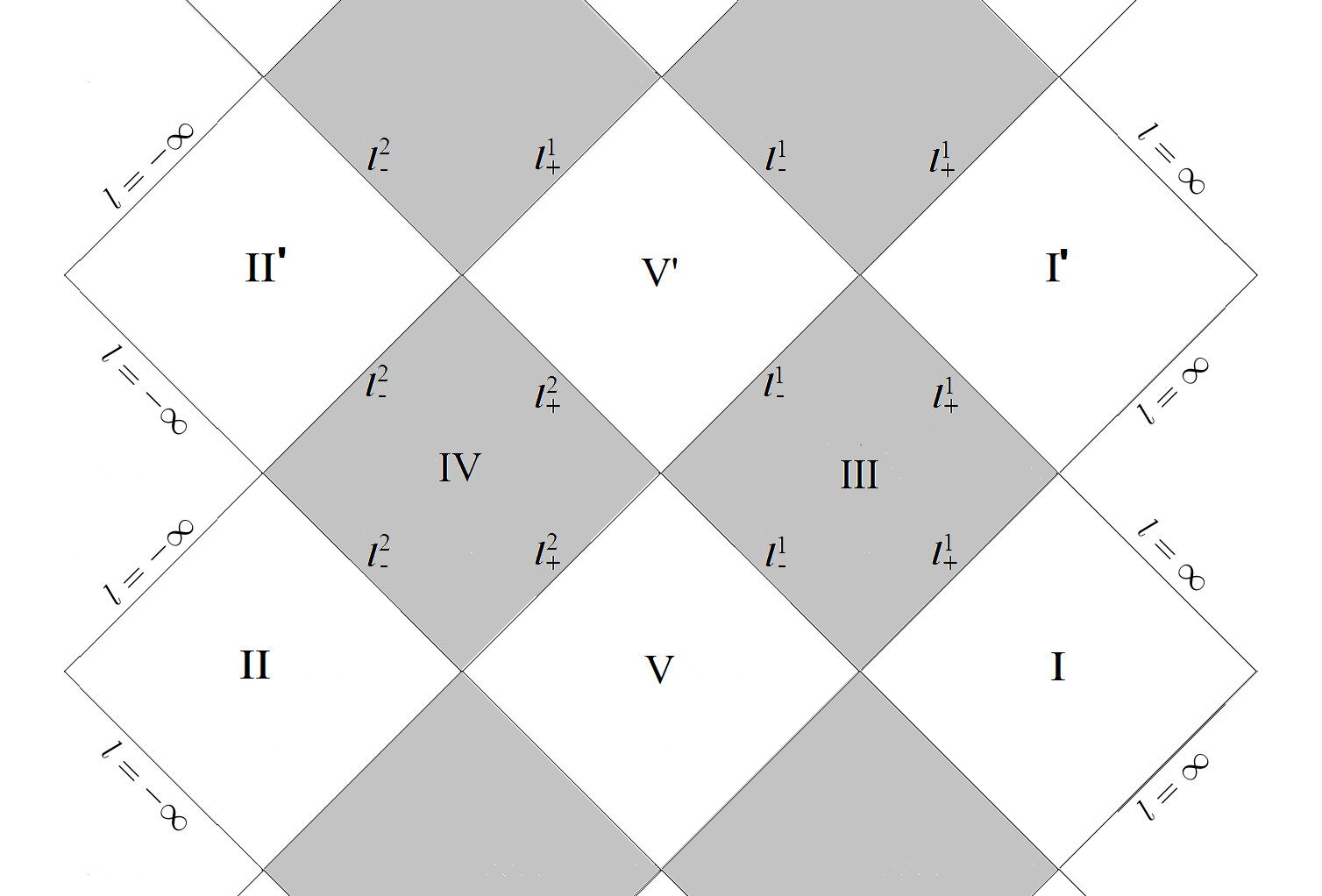}
  \caption{Penrose diagrams of the three possible causal structure found in the vacuum solutions of ansatz (\ref{ansatz}). The shaded regions have $G<0$ and the $l=const$ slices are spacelike. \bf Top: \rm  $G$ is well behaved and no horizons exist. \bf Middle: \rm $G$ is negative in a single region between $l\in[l_-,l_+]$.  \bf Bottom: \rm $G$ is negative in two regions between $l\in[l^1_-,l^1_+]$ and $l\in[l^2_-,l^2_+]$. }\label{horizons}
\end{figure}

For certain initial conditions our solutions can develop horizons, however the solutions remain free of singularities. To the asymptotic observer these solutions appear as non-singular black holes (or one-way traversable
wormholes). In general there can be up to two regions where the $l=const$ slices become spacelike, whenever $G$ becomes negative. This can be seen by noting that for $|l/k|>1$ the ODE is similar to an overdamped oscillator with equilibrium at $G=1$ and therefore any zero crossings must occur in the range $l\in [-k,k]$. As $\mathcal{O}(1)$ changes in $G$ occur on a length scale $l\sim k$, we expect $G$ to oscillate at most once during this interval, possibly below zero, and then it will asymptote monotonically to its equilibrium value. This may occur on either side of the $l=0$ point and  there are at most two regions where $G$ is negative. 
Penrose diagrams show the causal structures of the three spacetimes discussed above in Figure \ref{horizons}. In the remainder of this work we focus on the cases with no horizons, where $G$ remains well behaved so that the $\{t,l,\theta,\phi\}$ coordinate system covers the entire spacetime manifold. In other words we will assume that $G$ is a continuous non vanishing function of $l$ that asymptotes to 1 as $l\rightarrow \pm \infty$. 

 \section{Geodesics of Timelike and Null Observer}\label{sec_geoobs}
Here we compute the paths of null and timelike geodesics traveling in the radial direction ($d\Omega=0$). We show that a timelike observer can, in principle, safely traverse the wormhole. 
We begin with a discussion of null geodesic which have $ds^2=0$ and are therefore parametrized by,
 \be
 \frac{dl}{dt}=\pm G.
 \ee
 
Assuming no horizons, this may be integrated to find the path $l(t)$ of radially propagating photons. In section \ref{sec_div} we shall see that the affine parameter defined by $dt/d\lambda=c G^{-1}$ for some arbitrary constant $c$ is a monotonic boundless function of $t$. Hence the light rays trajectory through the throat can be described in a finite range of the affine parameter.

\begin{figure}[]
\centering
  \begin{subfigure}[]{0.35\textwidth}
  \includegraphics[width=\textwidth]{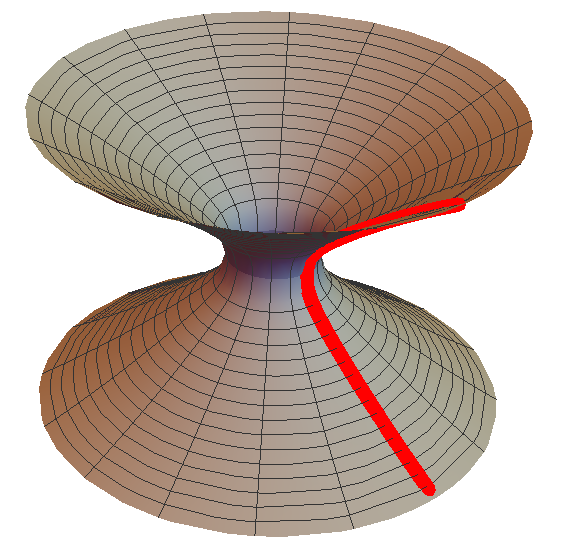}
 \caption{
Embedding of the wormhole in higher dimensions with a throat--traversing, radial trajectory traced in red. We have suppressed the $\theta$ coordinate such that every circle at some constant $z$ is a sphere having that circle's radius.}\label{wormgeo}
  \end{subfigure}
  \begin{subfigure}[]{0.4\textwidth}
  \includegraphics[width=\textwidth]{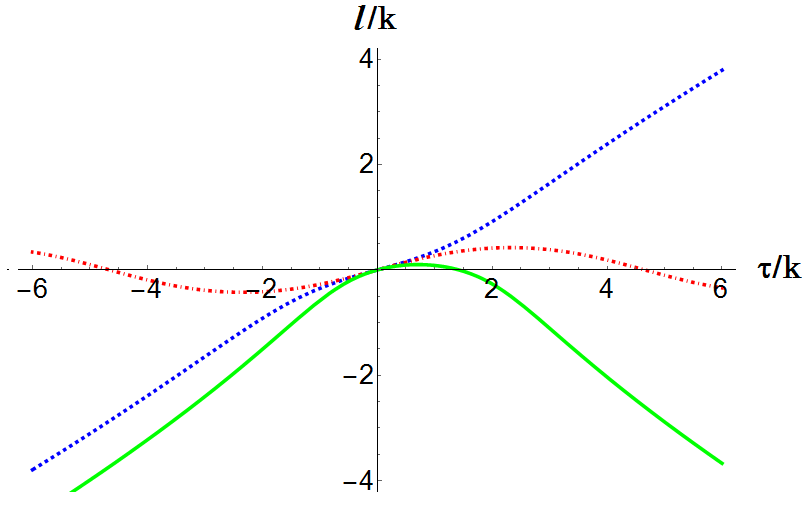}
  \includegraphics[width=\textwidth]{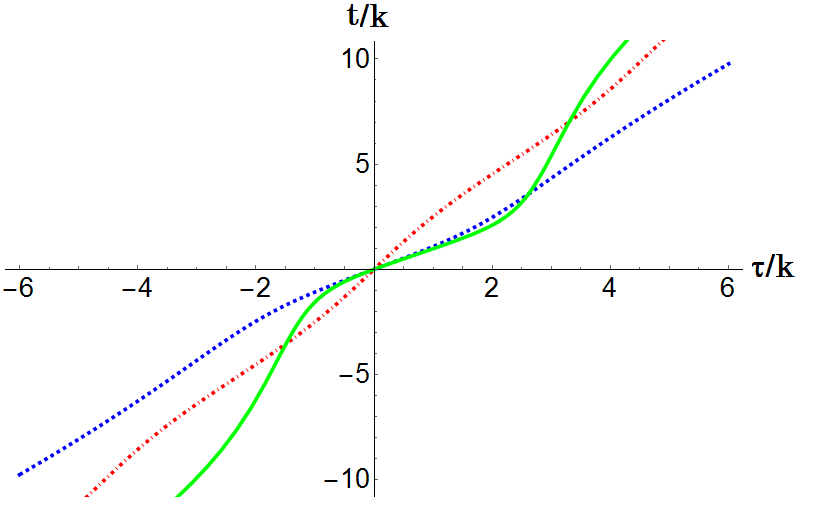}
  \caption{
The coordinates $l$ and $t$ as seen by observers with identical initial speeds at $l=0$ and attempting to cross the different throats. The observer on the green geodesic fails to traverse the wormhole and is spit back out. The observer labelled by the red dot-dashed line oscillates around the everywhere--attractive wormhole. The blue dashed observer successfully passes through the throat.}\label{ltvsPT}
  \end{subfigure}
  \caption{Wormhole structure.}
\end{figure}

We now turn to timelike geodesics.  Given a four velocity $u^\alpha(\tau)$ with $\tau$ being the observer's proper time the geodesic equation reads,
\be
u^\alpha\nabla_\alpha u^\beta =0.
\ee
The normalization of ${\mathbf{u}}$ for radial trajectories gives the following condition,
\be\label{normalization}
\Big(\frac{dt}{d\tau}\Big)^2=\frac{1}{G^2}\Big(\frac{dl}{d\tau}\Big)^2+\frac{1}{G}.
\ee
Eliminating $u^0=dt/d\tau$ from the $l$ component of the geodesic equation, one can solve for $l(\tau)$ through,
\be\label{eoml}
\frac{\text{d}^2}{\text{d}\tau^2}l=-\frac{G'}{2} \,,
\ee
and $t(\tau)$ is determined from Eq.(\ref{normalization}). To visualize the geodesic motion, we embed the geometry in a higher dimensional space. Suppressing the $\theta$ coordinate and taking a time slice at a particular $t$ (which looks the same for any choice of $t$), we embed the wormhole in a 3D space as shown in figure (\ref{wormgeo}). This particular solution corresponds to a $G_0=1$ and $v_0=0$ wormhole. The red line represents the radial trajectory of a timelike observer.
We also show how different values of $G_0,~v_0$ (defined and color coded in the caption of figure (\ref{worm1})) affect our courageous explorer along his journey.

Note that the solutions can have regions where gravity is attractive or repulsive towards $l=0$. The transition occurs wherever $G'$ switches sign. For instance, our solutions with $G_0=1$ have repulsive gravitational forces near the throat and incoming observers must possess sufficient velocity to traverse the throat. In figure (\ref{ltvsPT}), our infalling explorer labelled by the solid green curve does not have sufficient speed and is repelled out of the wormhole. However, the solution with $G_0=0.2$ is attractive everywhere towards the wormhole and the observer travelling in this background (designated by the red dot-dashed line in figure (\ref{ltvsPT})) oscillates back and forth through the wormhole. It is even easy to find solutions where $G$ asymptotically relaxes back to $1$ from above. In these cases our asymptotic observers would associate a negative $M_{\pm}$ to the wormhole and feel a repulsive force from it! This might be concerning at first sight as this would imply a spacetime with negative mass in standard GR. However the situation is more complicated in $R^2$ gravity as we will discuss in section \ref{sec_asymptoticsol}.
\section{Tidal forces through the wormhole}\label{sec_div}
In this section we determine how bundles of geodesics evolve as they pass through the wormhole. The non-zero components of the Ricci tensor read,
\begin{align}
R_{tt}&=\frac{1}{2}G\Big(\frac{2l G'}{k^2+l^2}+G''\Big),\\
R_{rr}&=-(2G)^{-1}\Big(\frac{4k^2G+2l(k^2+l^2)G'}{(k^2+l^2)^2}+G''\Big),\\
R_{\theta\theta}&=1-G-lG',~~~R_{\phi\phi}=\sin(\theta)^2 R_{\theta\theta}.
\end{align}
Taking the radial geodesics with affine parameter $\lambda$, ${\bf{k}}=\frac{\partial t}{\partial \lambda} (1,\pm G,0,0) $ for in(-) and outgoing(+) geodesics respectively, we find
\begin{align}
R_{\mu\nu}&k^\mu k^\nu=-\Big(\frac{\partial t}{\partial\lambda}\Big)^2\frac{2k^2 G^2}{(k^2+l^2)^2}.
\end{align}
This is strictly negative and therefore can allow for $\frac{\text{d}}{\text{d}\lambda}\theta>0$. We now compute the divergence explicitly. The divergence of a bundle of geodesics with tangent vectors $k^\mu$ is given by $\theta=\nabla_\alpha k^\alpha=\frac{1}{\sqrt{-g}}\partial_\alpha(\sqrt{-g} k^\alpha)$. To determine $k^t=\frac{\partial t}{\partial\lambda}$ one must look at the geodesic equation for the null ray. The $k^t$ equation is,
\begin{align}
k^t\frac{d}{dt}k^t+k^l\frac{d}{dl}k^t+\Gamma^t_{\alpha\beta}k^\alpha k^\beta&=0,\\
\implies ~~\frac{d}{dl}k^t+\frac{G'}{G}k^t&=0\,,
\end{align}
where in the second line we assumed $\frac{d}{dt}k^t=0$, appropriate for a time independent metric. This is solved by $k^t=c G^{-1}$ for some integration constant $c$ which denotes the arbitrary choice in the affine parameter's normalization. We take the value  $c=1$. In the cases were $G$ is well behaved so that our coordinate system is valid everywhere, we can compute $\theta$ for any $l$,
\be
\theta=\pm \Big(\frac{2l}{l^2+k^2}\Big).
\ee
The $\pm$ corresponds to in and outgoing modes respectively. As expected, $\theta$ switches sign when the light rays cross the throat at $l=0$. Moreover, since $\theta$ gives the expansion/contraction of the cross sectional area of a bundle of light rays (it is an area, as orthogonal hypersurfaces to null geodesics are two dimensional), the radial null geodesics of the metric in Eq. (\ref{ansatz}) are not affected by the form of $G$.

A timelike observer will also feel increasingly squeezed by the tidal forces as we enters the wormhole. Unlike for null geodesics, such observers will be sensitive to $G$. Morris \& Thorne \cite{Morris:1988cz} used these tidal forces to impose constraints on the wormhole spacetime by requiring that a human sized observer feels less than a $g$ of force.  
We avoid imposing any such conditions here but give a brief derivation of the tidal forces below. 
We refer the reader to \cite{Morris:1988cz} for additional details.

Consider an observer $\mathcal{O}'$ at rest with respect to $\{l,\theta,\phi\}$, and proper reference frame with basis vectors,
\begin{align}
{\mathbf{e}}_{\hat{t}}&=\frac{1}{\sqrt{G}}{\mathbf{e}}_{t}, ~{\mathbf{e}}_{\hat{l}}=\sqrt{G}{\mathbf{e}}_{l},\nonumber\\
{\mathbf{e}}_{\hat{\theta}}&=\frac{1}{\sqrt{k^2+l^2}}{\mathbf{e}}_{\theta},~{\mathbf{e}}_{\hat{\phi}}=\frac{1}{\sin\theta\sqrt{k^2+l^2}}{\mathbf{e}}_{\phi}.
\end{align}
In this coordinate system the metric takes the simple form $\eta_{\hat{\alpha}\hat{\beta}}=(-1,1,1,1)$.
A second observer $\mathcal{O}''$ located at the same point, but moving with a speed $v$ in the radial direction, will also have his own reference frame whose basis can be written in terms of the coordinates of $\mathcal{O}'$ through a Lorentz boost, leaving $\eta_{\hat{\alpha}\hat{\beta}}$ unchanged,
\begin{align}
{\mathbf{e}}_{\hat{0}}&=\gamma{\mathbf{e}}_{\hat{t}}+ \gamma v{\mathbf{e}}_{\hat{l}},\nonumber\\ ~{\mathbf{e}}_{\hat{1}}&=\gamma v{\mathbf{e}}_{\hat{t}}+ \gamma{\mathbf{e}}_{\hat{l}},\nonumber\\
{\mathbf{e}}_{\hat{2}}&={\mathbf{e}}_{\hat{\theta}},~{\mathbf{e}}_{\hat{3}}={\mathbf{e}}_{\hat{\phi}}.
\end{align}
Here the gamma factor is the usual $\gamma=(1-v^2)^{-1/2}$. In this frame the moving observer has four velocity  ${\mathbf{\tilde{u}}}={\mathbf{e}}_{\hat{0}}$. Since the four acceleration is perpendicular to the four velocity, our radially moving observer's acceleration will only have a non zero radial component in his own reference frame, namely ${\mathbf{a}}=a {\mathbf{e}}_{\hat{1}}$. To find $a$ is to write ${\mathbf{a}}=\frac{d}{d\tau}{\mathbf{\tilde{u}}}$ in the $\{t,l,\theta,\phi\}$ coordinates which have constant vielbeins and use,
\begin{equation}
\tilde{u}^\alpha \nabla_\alpha \tilde{u}_{t}={\mathbf{a}}\cdot {\mathbf{e}}_t=a{\mathbf{e}}_{\hat{1}}\cdot{\mathbf{e}}_t \,.
\end{equation}
Here, $\tilde{u}_t={\mathbf{e}}_{\hat{0}}\cdot{\mathbf{e}}_{t}$, and,
\be\label{acceleration}
a=-\frac{1}{\sqrt{G}}\Big[\gamma\frac{G'}{2}+G\frac{\partial \gamma}{\partial l} \Big].
\ee
At $l\rightarrow \pm \infty$ and assuming $\gamma\approx 1$, the acceleration felt is $a\approx -M_\pm/l^2$. For $l\approx 0$, the term $G'$ contains terms evolving as $G_0/k$ and $v_0$. Hence one may tune these values so that the observer crossing the throat will feel a desired acceleration.

The tidal acceleration felt by the observer $\mathcal{O}''$ can also be computed in his reference frame. The relative acceleration of two nearby points separated by ${\mathbf{\xi}}=(0,\vec{\xi})$ as measured by $\mathcal{O}''$ is given by,
\be
\Delta a^{\hat{j}}=-R^{\hat{j}}_{\hat{\alpha}\hat{\beta}\hat{\rho}}u^{\hat{\alpha}}\xi^{\hat{\beta}}u^{\hat{\rho}}=- R_{\hat{j}\hat{0}\hat{k}\hat{0}}\xi^{\hat{k}},
\ee
with,
\begin{align}
R_{\hat{1}\hat{0}\hat{1}\hat{0}}&=-\frac{G''}{2},~~~~R_{\hat{2}\hat{0}\hat{2}\hat{0}}=R_{\hat{3}\hat{0}\hat{3}\hat{0}},\nonumber \\
R_{\hat{3}\hat{0}\hat{3}\hat{0}}&=\gamma^2 \frac{lG'}{l^2+k^2}-\gamma^2v^2\Big(\frac{l(l^2+k^2)G'+2k^2G}{2(l^2+k^2)^2}\Big).
\end{align}
We can explicitly see how the size of the throat $k$ influences the strength at which our traveler is squeezed in the angular directions.

\section{The wormhole to asymptotic observers}\label{sec_asymptoticsol}
From Figure (\ref{worm1}) we see that there is a lack of parity symmetry in the $l$ component when $v_0\ne 0$. As previously mentioned, this causes observers in different asymptotic regions to see different looking wormholes, characterized by $M_\pm$ and $Q_\pm$ seen at $l=\pm\infty$. These are determined by $G_0,~v_0$ from expanding the exact solution of $G$ and picking out the asymptotic coefficients. Defining $u=1/l$, we expand the solution of $G$ around $u=0$ and define,

\begin{align}
M_\pm&=\lim_{u\rightarrow\pm 0} \frac{-1}{2}\frac{\text{d}}{\text{d}u}G(u,G_0,v_0),\label{Mdef}\\
Q_\pm^2&=\lim_{u\rightarrow\pm 0} \frac{\text{d}^2}{\text{d}u^2}G(u,G_0,v_0).\label{Qdef}
\end{align}

 One finds that,
\be
\frac{\mp 2M_\pm}{k}=-2\frac{M_0}{k}+\cos(\sqrt{3} \pi/2) G_0\pm \frac{\sin (\sqrt{3} \pi/2)}{\sqrt{3}}kv_0 \,
\ee
with $M_0\approx 0.7$ being a number whose exact expression is given in Appendix (\ref{appA}). Note that if $G$ is to approach one from below in the $l\rightarrow -\infty$ limit, then the sign of $M_-$ will be negative but this is just because our "radial" coordinate $l$ takes on negative values. The $\pm$ sign in front of $v_0$ is responsible for the asymmetry. Even though this implies that asymptotic observers will feel different accelerations toward the wormhole, one cannot gain energy by going from one region to the other. The asymmetrical shape of the throat (as is seen by the solid green curve in Fig.\ref{worm1}) will prohibit this as can be shown by Eq.(\ref{eoml}) which implies $(dl/d\tau)^2+G$ is a conserved quantity and so $\Delta v =[dl/d\tau]_{l=\pm\infty}=0$.

Meanwhile the coefficient $Q_\pm^2$ is found to be,
\be
\frac{Q_\pm^2}{k^2}=\frac{Q_0^2}{k^2}\pm\sqrt{3}\sin (\sqrt{3} \pi/2)G_0-  \cos(\sqrt{3} \pi/2)kv_0-2\log (u\text{k}).
\ee

The very last term has a logarithmic dependence so that, not only is $Q_\pm^2$ not a constant, it diverges as $u\rightarrow 0$. This shows that the exact solution and the Reissner\textendash Nordstr{\"o}m spacetime differs radically in the asymptotic region. In the case of GR with electromagnetism, the conserved charge is defined by the conserved current $j^\mu=\frac{1}{4\pi}\nabla_\nu F^{\mu\nu}$ integrated over a spacelike hypersurface. Here no such analogue electromagnetic field is present. The logarithm found is just a feature of our particular wormhole solution. Choosing a different ansatz, such as simply replacing the angular part of Eq. (\ref{ansatz}) by,
\be
g_{\Omega}\text{d}\Omega^2=\Big(l^2+\frac{k^2}{l^2+k^2}\Big)\text{d}\Omega^2,
\ee
allows us to obtain wormhole solutions that asymptotically approach Reissner\textendash Nordstr{\"o}m metrics without logarithmic contribution.

Conserved charges are tightly linked with symmetries of the theory. In our case, we have a timelike killing vector ${\bf{e}}_{t}=\frac{\partial}{\partial t}$ with which we can construct a conserved current and find the corresponding charge; however, the interpretation of charges in $R^2$ gravity is tricky. Previous attempts \cite{Deser:2002rt, Deser:2002jk} by Deser \& Tekin, to define an energy for higher curvature gravity in analogy with GR have shown that every asymptotically flat solution of $R^2$ gravity has vanishing energy. An alternative definition was proposed by the same authors and relied instead on the corresponding Poisson equation of higher curvature gravity \cite{Deser:2007vs}. In terms of this second definition, the leading term of the source in the Poisson equation  defines the energy and allows for a less degenerate classification of spacetimes but still yields a vanishing energy for solutions with everywhere vanishing Ricci scalar. Therefore even though an asymptotic observer will feel that the wormhole is exerting a gravitational force (just as if it were a spherical object of mass $M_\pm$ in ordinary GR) he shouldn't associate this value with a conserved charge such as the energy. Hence one might wonder if these quantities can evolve upon the introduction of small perturbations.

\section{Conclusion}
In this paper we have presented new static wormhole and (non-singular) black hole solutions found in the vacuum of $R^2$ gravity. The traversable wormholes are are solutions in $D = 4$ which do not require the support of any NEC violating exotic matter. Their existence could prove important as it is argued in \cite{Kehagias:2015ata} that the new $R=0$ solutions might be part of the strong coupling limit ($M_p\rightarrow 0$) of General Relativity. It would be interesting to study the role of such spacetimes in this limit using other tools. For instance it was shown recently that in the context of AdS/CFT, the strong subadditivity condition of entanglement entropy imposes constraints on the bulk geometry of certain spacetimes \cite{Lashkari:2014kda}. One would require a dual CFT to asymptotically $R=0$ solutions to apply similar results.

In the context of our particular solution, we have shown that asymptotic observers view the wormhole as a black hole with ``mass" $M_\pm$ and varying ``charge" $Q_\pm$. We have stressed that one should not be associating these values as conserved charge. They are taken as coefficients of the $1/r$ and $1/r^2$ terms in the metric expansion and do not correspond to some conserved quantity as was the case in GR. This raises the question of whether $M_\pm$ and $Q_\pm$, along with the whole wormhole itself, are dynamical once small perturbations are introduced. It would also be interesting to look for further exotic solutions to general $f(R)$ theories which cannot be captured by Einstein gravity coupled to a scalar field.
\section*{Acknowledgements}
We are grateful to Alex Belin, Eliot Hijano, Maulik Parikh, and G. Salton for helpful discussions. 
FD is supported by a B2 scholarship from le Fonds de recherche du Qu\'ebec\textendash Nature et technologies (FRQNT).

\begin{widetext}
\appendix
\section{}\label{appA}

The general solution to $G$ is given by
\begin{align}
G(l)=&1+\frac{ k \cos \big(\sqrt{3} \arctan(l/k)\big)}{\sqrt{k^2+l^2}}G_0+\frac{k \sin \big(\sqrt{3} \arctan(l/k)\big)}{\sqrt{3} \sqrt{k^2+l^2}}kv_0\nonumber \\
&-\frac{k}{6 \sqrt{k^2+l^2}}\Big[  \big(\sqrt{3} H_{-\frac{1}{4}(1+\sqrt{3})}-\sqrt{3} H_{\frac{1}{4} (-5+\sqrt{3})}-2 \sqrt{3}-3+\sqrt{3} \pi  \tan \big((1+\sqrt{3})\pi/4 \big)\big) e^{ i \sqrt{3} \arctan(l/k)}\nonumber \\
&~~~~~~~~~~~~~~~~~~~~~~~~~~~~~~~~~~~~~~~~~~~~~~~~~~~+\big(\sqrt{3} H_{\frac{1}{4} (-3+\sqrt{3})}-\sqrt{3} H_{\frac{1}{4}(-1+\sqrt{3})}+3\big)e^{-i \sqrt{3} \arctan(l/k)}\Big]\nonumber \\
&-\frac{2k}{(3+\sqrt{3}) (k-i l)}\Big[(2+\sqrt{3})  \, _2F_1\big(1,\frac{1}{2}(1-\sqrt{3});\frac{1}{2} (3-\sqrt{3});1-\frac{2 k}{k-i l}\big)\nonumber\\
&~~~~~~~~~~~~~~~~~~~~~~~~~~~~~~~~~~~~~~~~~~~~~~~~~~~~~~~~+  \, _2F_1\big(1,\frac{1}{2} (1+\sqrt{3});\frac{1}{2} (3+\sqrt{3});1-\frac{2 k}{k-i l}\big)\Big].
\end{align}
Here $_2F_1(a,b;c;z)$ is the ordinary hypergeometric function.  The coefficients $M_\pm$ and $Q_\pm$ can be read from the series expansion of $G$ at large $l$ as written in Eq. (\ref{Mdef}, ~\ref{Qdef}). The result is,
\begin{align}
\frac{\mp 2M_\pm}{k}=&\cos(\sqrt{3} \pi/2)G_0\pm \frac{\sin (\sqrt{3} \pi/2)}{\sqrt{3}}kv_0\nonumber\\
&+\frac{1}{2 \sqrt{3}}\cos(\sqrt{3} \pi/2) \left[-H_{-\frac{1}{4} (1+\sqrt{3})}+2 H_{\frac{1}{4} (-5+\sqrt{3})}-H_{-\frac{1}{4} (3-\sqrt{3})}+2 \sqrt{3}+4-\pi  \tan\big( (1+\sqrt{3}) \pi/4 \big)\right],\\
\frac{Q^2}{k^2}=&\pm\sqrt{3} \sin (\sqrt{3} \pi/2)G_0-  \cos(\sqrt{3} \pi/2)kv_0\nonumber\\
&+\frac{1}{2} \sin (\sqrt{3} \pi/2) \left[-H_{-\frac{1}{4} (1+\sqrt{3})}+2 H_{\frac{1}{4} (-5+\sqrt{3})}-H_{-\frac{1}{4} (3-\sqrt{3})}+2 \sqrt{3}+4-\pi  \tan\big( (1+\sqrt{3}) \pi/4 \big)\right]\nonumber\\
&-2 \log (2u\text{k})-2(1-\gamma_{E-M})-\psi ^{(0)}\left((1-\sqrt{3})/2\right)-\psi ^{(0)}\left((1+\sqrt{3})/2\right)
\end{align}
Here $\gamma_{E-M}$ is the Euler-Mascheroni constant. The functions $H_n$ and $\psi ^{(0)}(z)$ are Harmonic numbers and polygamma functions respectively, both are defined from the Gamma function through,
\begin{align}
H_n&=\gamma_{E-M}+\frac{\Gamma'(n+1)}{\Gamma(n+1)},\\
\psi ^{(0)}(z)&=\Gamma'(z)/\Gamma(z).
\end{align}
\end{widetext}


\begin{thebibliography}{10}
\bibitem{Kehagias:2015ata} 
  A.~Kehagias, C.~Kounnas, D.~Lust and A.~Riotto,
  arXiv:1502.04192 [hep-th].
\bibitem{Stelle:1976gc} 
    K.~S.~Stelle,
    Phys.\ Rev.\ D {\bf 16}, 953 (1977).
\bibitem{Lu:2015cqa} 
  H.~Lu, A.~Perkins, C.~N.~Pope and K.~S.~Stelle,
  Phys.\ Rev.\ Lett.\  {\bf 114}, no. 17, 171601 (2015)
  [arXiv:1502.01028 [hep-th]].
\bibitem{Kounnas:2014gda} 
  C.~Kounnas, D.~Lust and N.~Toumbas,
  Fortsch.\ Phys.\  {\bf 63}, 12 (2015)
  [arXiv:1409.7076 [hep-th]].
\bibitem{Alvarez-Gaume:2015rwa} 
  L.~Alvarez-Gaume, A.~Kehagias, C.~Kounnas, D.~Lust and A.~Riotto,
  arXiv:1505.07657 [hep-th].
\bibitem{Morris:1988cz} 
  M.~S.~Morris and K.~S.~Thorne,
  Am.\ J.\ Phys.\  {\bf 56}, 395 (1988).
\bibitem{James:2015ima} 
  O.~James, E.~von Tunzelmann, P.~Franklin and K.~S.~Thorne,
  arXiv:1502.03809 [gr-qc].
\bibitem{VanRaamsdonk:2010pw} 
  M.~Van Raamsdonk,
  Gen.\ Rel.\ Grav.\  {\bf 42}, 2323 (2010)
  [Int.\ J.\ Mod.\ Phys.\ D {\bf 19}, 2429 (2010)]
  [arXiv:1005.3035 [hep-th]].
\bibitem{Maldacena:2013xja} 
  J.~Maldacena and L.~Susskind,
  Fortsch.\ Phys.\  {\bf 61}, 781 (2013)
  [arXiv:1306.0533 [hep-th]].
  
\bibitem{FMM} V. Frolov, M. Markov and V. Mukhanov {\it Phys. Lett.} {\bf B216}, 272 (1989).

\bibitem{PBB} G. Veneziano, {\it Phys. Lett.} {\bf B265}, 287 (1991).

\bibitem{Morgan} D. Morgan, {\it Phys. Rev.} {\bf D45}, R1005 (1992).

\bibitem{MB} V. Mukhanov and R. Brandenberger, {\it Phys. Rev. Lett. } {\bf 68}, 1969 (1992).

\bibitem{Smolin}  L. Smolin, {\it Class. Quant. Grav.} {\bf 9}, 173 (1992).

\bibitem{TV} A. Tseytlin and C. Vafa, {\it Nucl. Phys.} {\bf B372}, 443 (1992).

\bibitem{TMB} M. Trodden, V. Mukhanov and R. Brandenberger, {\it Phys. Lett. } {\bf B316}, 483 (1993).

\bibitem{Lowe} G. Alberghi, D. Lowe and M. Trodden, {\it JHEP} {\bf 9907:020 }(1999).

\bibitem{DKH} R. Daghigh, J. Kapusta and Y. Hosotani, ``False Vacuum Black Holes and Universes'',
hep-ph/0008006, (2000).

\bibitem{Easson:2001qf} 
  D.~A.~Easson and R.~H.~Brandenberger,
  JHEP {\bf 0106}, 024 (2001)
  [hep-th/0103019].
\bibitem{Easson:2002tg} 
  D.~A.~Easson,
  JHEP {\bf 0302}, 037 (2003)
  [hep-th/0210016].
\bibitem{Hochberg:1990is} 
  D.~Hochberg,
  Phys.\ Lett.\ B {\bf 251}, 349 (1990).
\bibitem{Furey:2004rq} 
  N.~Furey and A.~DeBenedictis,
  Class.\ Quant.\ Grav.\  {\bf 22}, 313 (2005)
  [gr-qc/0410088].
\bibitem{Lobo:2009ip} 
  F.~S.~N.~Lobo and M.~A.~Oliveira,
  Phys.\ Rev.\ D {\bf 80}, 104012 (2009)
  [arXiv:0909.5539 [gr-qc]].
\bibitem{Oliveira:2011vu} 
  M.~A.~Oliveira,
  arXiv:1107.2703 [gr-qc].
\bibitem{DeBenedictis:2012qz} 
  A.~DeBenedictis and D.~Horvat,
  Gen.\ Rel.\ Grav.\  {\bf 44}, 2711 (2012)
  [arXiv:1111.3704 [gr-qc]].
\bibitem{Harko:2013yb} 
  T.~Harko, F.~S.~N.~Lobo, M.~K.~Mak and S.~V.~Sushkov,
  Phys.\ Rev.\ D {\bf 87}, no. 6, 067504 (2013)
  [arXiv:1301.6878 [gr-qc]].
\bibitem{DeFelice:2010aj} 
  A.~De Felice and S.~Tsujikawa,
  Living Rev.\ Rel.\  {\bf 13}, 3 (2010)
  [arXiv:1002.4928 [gr-qc]].
\bibitem{Deser:2002rt} 
  S.~Deser and B.~Tekin,
  Phys.\ Rev.\ Lett.\  {\bf 89}, 101101 (2002)
  [hep-th/0205318].
\bibitem{Deser:2002jk} 
  S.~Deser and B.~Tekin,
  Phys.\ Rev.\ D {\bf 67}, 084009 (2003)
  [hep-th/0212292].
  \bibitem{Deser:2007vs} 
    S.~Deser and B.~Tekin,
    Phys.\ Rev.\ D {\bf 75}, 084032 (2007)
    [gr-qc/0701140].
\bibitem{Lashkari:2014kda} 
  N.~Lashkari, C.~Rabideau, P.~Sabella-Garnier and M.~Van Raamsdonk,
  arXiv:1412.3514 [hep-th].

\end{thebibliography}
\end{document}